\documentclass[reqno,letterpaper, 12pt]{amsart}

\usepackage{graphicx}
\usepackage[left=1.in,right=.75in,top=1.in,bottom=1.in,
            headheight=18pt,headsep=18pt,footskip=24pt]{geometry}

\usepackage{amsmath}

\newcommand{\abs}[1]{\left\vert#1\right\vert}

\begin{document}
\title[Numerical evaluation of relative permeability]%
{Numerical evaluation of relative permeability using Johnson--Koplik--Dashen model}%
\author{Andrea Cortis}%
\address{ION Geophysical, 2105 Citywest blvd, Suite 900, Houston, TX, 77042, USA}%
\email{andrea.cortis@iongeo.com}%
\author{Dmitriy Silin}%
\address{Shell Corporation}%
\email{dbsberkeley@gmail.com}%

\date{\today}%

\begin{abstract}
We present a numerical study aimed at comparing two approaches to the evaluation of relative permeability curves from 3D binary images of porous media.  
One approach hinges on the numerical solution of Stokes equations, while the other is based on the Johnson-Koplik-Dashen (JKD) universal scaling theory of viscous frequency-dependent flow [D.~L. Johnson, J.~Koplik, and R.~Dashen, \emph{Theory of dynamic permeability and tortuosity in fluid--saturated porous media}, Journal of Fluid Mechanics \textbf{176} (1987), 379--402.] and 
the method of maximal inscribed spheres.
JKD steady-flow simulations only require the solution of a boundary-value problem for the Laplace equation, which is computationally less intensive than the solution of Stokes equations. 
A series of numerical calculations performed on 3D pore-space images of natural rock demonstrate that JKD-based estimates are in good agreement with the corresponding Stokes-flow numerical simulations.
\end{abstract}

\maketitle
\section{Introduction}


Consider a porous medium whose pore space is occupied by two immiscible fluid phases under capillary equilibrium moving under the action of their own pressure gradients.
Assuming quasi-static flow, at any given saturation $s$ (i.e., the relative volume of the wetting phase)
the fluid phases distribution is not significantly different than the one for static capillary equilibrium, and we can write the classical set of equations \cite[]{MuskatMeresPhysics:1936}

\begin{equation}\label{eq:relative_permeability}
    v_i = -k \frac{k_{ri}}{\mu_i} \nabla p_i, \quad i=1,2 \, ,
\end{equation}
where $k$ is the rock permeability, $v_i$, $\mu_i$, $p_i$, and $k_{ri}$ are the Darcy velocity, viscosity, pressure, and relative permeability of fluid phase $i$, respectively.

The experimental evaluation of the macroscopic parameters involved in two-phase flow equations is often plagued by error and uncertainty.
Alternatively, relative permeability curves can be estimated numerically.
While numerical simulations cannot replace the experiment, they may help to reduce the uncertainties of interpretation of the results.
It is well known that the evaluation of Darcy permeability, $k$, for a given porous geometry can be obtained by averaging single-phase creeping flow velocity (i.e., Reynolds number $Re \sim 0$) assuming no-slip boundary conditions at the fluid-rock interface~\cite[]{Hubbert:1940}.

The numerical evaluation of the relative permeability can be similarly obtained when the precise distribution of the two phases  is known for a given saturation.
A no-slip boundary condition characterizes the interaction of fluid and solid whereas the determination of the fluid-fluid interaction is less certain and depends, for instance, on the local co-current and counter-current flows. In this work, as a first approximation, we impose no-slip boundary conditions on the fluid-fluid interfaces and leave the investigation of possibly more 
adequate 
boundary conditions to future studies.
Stokes flow equation are then solved in the pore space portion occupied by the mobile phase estimated with a MIS algorithm~\cite[]{MISVerification:2010}.
The resulting velocity field is averaged to obtain the Darcy velocity of the corresponding fluid phase and to estimate the value of relative permeability for the corresponding saturation.

Three-dimensional pore space geometry of rock can be described by a binary image obtained using high-resolution computed tomography (CT).   
The number of voxels necessary to characterize the micro-meter scale features of the pore space geometry can be of on the order of ${(10^4)}^3$. 
For practical applications,  numerical solution of the Stokes equations on a domain with so many grid cells can by computationally 
expensive. Alternatively, the permeability can be evaluated using the approach suggested by Johnson, Koplik, and Dashen (JKD) \cite[]{JKD87}.
This approach requires the solution of a single boundary-value problem for potential flow (Laplace equation), which is dramatically less computationally intensive than the solution of the viscous flow equations.
The JKD approach, however, involves an \textit{a priori} unknown proportionality factor $M$, which depends on the details of the pore space geometry (see eq.~\eqref{eq:M} below). If this factor remains constant for all saturations, it eventually cancels out in the estimation of the relative permeability curves for the given sample. Although, theoretically, the range of variation of $M$ is indefinite, in many practical situations it remains close to unity~\cite[]{JKD87,BJ87,CKS88,CORBOOK}.
It is to be noted here that the value of $M$ for each saturation is not merely a \textit{fitting parameter}, but it can actually be independently measured in a laboratory setting, or independently calculated. 
In this work, we will check
 numerically the consistency of these assumptions,
but we will not attempt at independent calculations of the $M$.

The general procedure proposed in the present work can be summarized as follows: given a 
binary image 
of a dry porous medium, 
we (\emph{i}) estimate the distribution of the wetting and nonwetting phases%
we (\emph{ii-a}) numerically estimate the relative permeability to each phase by solving the Stokes equations on the MIS-derived domains;  
we (\emph{ii-b}) numerically estimate the relative permeability to each phase by by using the JKD approximation on the MIS-derived domains;
and finally (\emph{iii}), we make an \textit{a posteriori} analysis of the proportionality factors $M$.
The  method of maximal inscribed spheres (MIS)~\cite[]{SilinPatzek:2006PhysicaA} was used to calculate the portion of the pore space occupied by each fluid and evaluate fluid saturation at a given capillary pressure.  In~\cite[]{MISVerification:2010}, the equilibrium two-phase fluid distribution computed with MIS was verified against experimental micro-tomography data, and it was obtained that a MIS-calculated capillary pressure curve can be in agreement with mercury porosimetry laboratory data.

\section{Frequency-dependent flow and permeability estimates}
The theory of dynamic permeability studies the flow of a viscous fluid inside a porous medium in response to a small-amplitude oscillatory changes in 
the pressure drop~\cite[]{LL59,JKD87}.
The essential idea behind this theory is that, as the frequency $\omega$ of the applied pressure drop oscillations increases,
the region of the pore space where the viscous dissipation occurs becomes localized to a narrow layer of thickness
$\delta \sim 1/\sqrt{\omega}$ adjacent to the fluid-rock interface ~\cite[]{LL59}.  
Outside this viscous dissipation layer, potential fluid flow is the dominant mechanism.
Following the developments in~\cite[]{JKD87}, it is possible to show that in a slab of porous material of thickness $L$ and unit cross-section bulk area
the high-frequency asymptotic expression for the real part of the complex-valued frequency-dependent dynamic permeability, $k(\omega)$, scales as $Re[k(\omega)] \sim \tfrac{1}{2}\sqrt{M}\omega^{-\frac{3}{2}}$,
where $M$ is a nondimensional coefficient defined as,
\begin{equation} \label{eq:M}
 M = \frac{8\alpha_\infty k}{\phi\Lambda^2}.  	
\end{equation}
Here
$\phi$ is the porosity and the quantities $\Lambda$ and $\alpha_\infty$ are defined as
\begin{align}
	\alpha_\infty &=\frac{\phi A \abs{\psi_L}^2}{
		L\int\abs{\mathbf{u}_p(\mathbf{r})}^2\,dV}
		\label{eq:alpha_inf} \\
	\Lambda &= 2\frac{\int\abs{\mathbf{u}_p(\mathbf{r})}^2\,dV}{
			\int\abs{\mathbf{u}_p(\mathbf{r})}^2\,dA}
			\label{eq:Lambda}	
\end{align}
see~\cite[Equations~(2.9) and~(2.17)]{JKD87}. The quantity $\alpha_\infty$ is also called \textit{tortuosity} \cite[]{BRO80}.
It can be demonstrated that its magnitude is greater or equal to one and the equality takes place only if the pore space is a bundle of straight channels.
In eqs.~\eqref{eq:alpha_inf} and~\eqref{eq:Lambda}, $\mathbf{u}_p(\mathbf{r})$ is the anti-gradient of the potential:
\begin{equation}
	\mathbf{u}_p(\mathbf{r}) = -\nabla\psi
\label{eq:nabla_psi}
\end{equation}
Here, $\psi$ is the solution to Laplace equation subject to boundary conditions $\psi=\psi_L$ at the outlet, $\psi=0$ at the inlet boundary, and the potential drop between the inlet and outlet boundaries is equal to  $\psi_L$.
At the pore walls inside the medium, the potential $\psi$ satisfies Neumann condition ${\partial\psi}/{\partial \mathbf{n}}=0$, where $\mathbf{n}$ is the surface normal. JKD permeability calculations
based on Eq.~\eqref{eq:M}
 are thus reduced to the evaluation of volume and surface integrals of a potential flow velocity.

Under the assumption of smooth pore geometry, one can therefore estimate the absolute permeability to each phase for any given saturation, $k'_{ai}(s)$ as:
\begin{equation}
  k'_{ai}(s) = M(s) \frac{\phi s \Lambda(s)^2}{8\alpha_\infty(s)}
\label{eq:JKDPerm_abs}
\end{equation}
where the dependence on saturation $s$ for all the potential flow parameters has been made explicit (see~\cite[Equation~(2.22)]{JKD87}.) 

The magnitude of the proportionality factor $M$ is close to unity for ``smooth'' pore geometries \cite[]{JKD87,BJ87,CKS88,CORBOOK}.
The small range of variation of the coefficient $M$ for smooth pore geometries is an indication of a universal scaling for the high-frequency behavior of the dynamic permeability \cite[]{JKD87}.
Significant deviations from this universal scaling behavior have been proved to arise in sharp-edge pore geometries~\cite[]{COR99.2,CSGL01} and for fractal channels~\cite{CORBER10}.

In our numerical simulations (see next Section), we will assume that, for the given pore geometry, $M$ does not depend on the value of the saturation, \textit{i.e.}, 
$M(s)=M(1)$   for all feasible $s$.  Here, saturation equal to one means that the respective quantity is evaluated for single-phase flow.
Thus, 
the JKD estimate of the relative permeability to each phase, $k'_{ri}(s) \equiv {k'_{ai}(s)}/{k'_{ai}(1)}$, can be calculated as
%
\begin{equation}
  k'_{ri}(s) = \frac{\frac{\phi s \Lambda(s)^2}{8\alpha_\infty(s)}}{\frac{\phi \Lambda(1)^2}{8\alpha_\infty(1)}} =
   s \left( \frac{\Lambda(s)}{\Lambda(1)} \right)^2\frac{\alpha_\infty(1)}{\alpha_\infty(s)}
\label{eq:JKDPerm}
\end{equation}

The condition $k'_{ri}(s)<1$ must be satisfied for every value of saturation $s$, i.e., 
\begin{equation}
   s \frac{\alpha_\infty(1)}{\alpha_\infty(s)} < \left( \frac{\Lambda(1)}{\Lambda(s)} \right)^2
\label{eq:condition}
\end{equation}

\section{From a voxel space to a tetrahedral mesh}
In this section, we describe  calculation of the relative permeability curves for two segmented 3D  binary images of sandstone samples (cases A and B).  The voxel size is in both cases 4.5~$\mu$m.

\begin{figure}[h]
\begin{center}
\includegraphics[width=12cm]{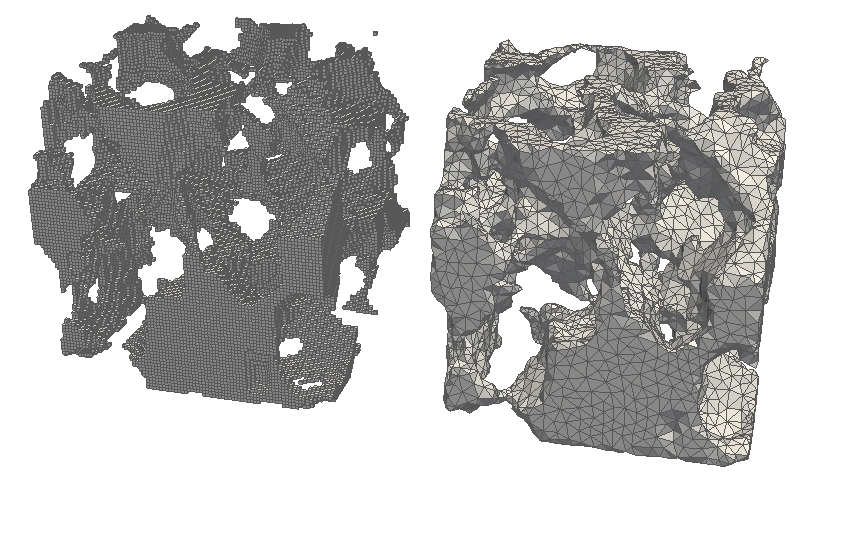}
\end{center}
\caption[]{Pore space corresponding to 0.40 mm linear size sandstone rock sample at saturation $s=0$: $90^3$ voxel segmentation of the original CT scan (left) and its tetrahedralization (Right).}
\label{fig:case_A}
\end{figure}

\begin{figure}[h]
\begin{center}
\includegraphics[width=12cm]{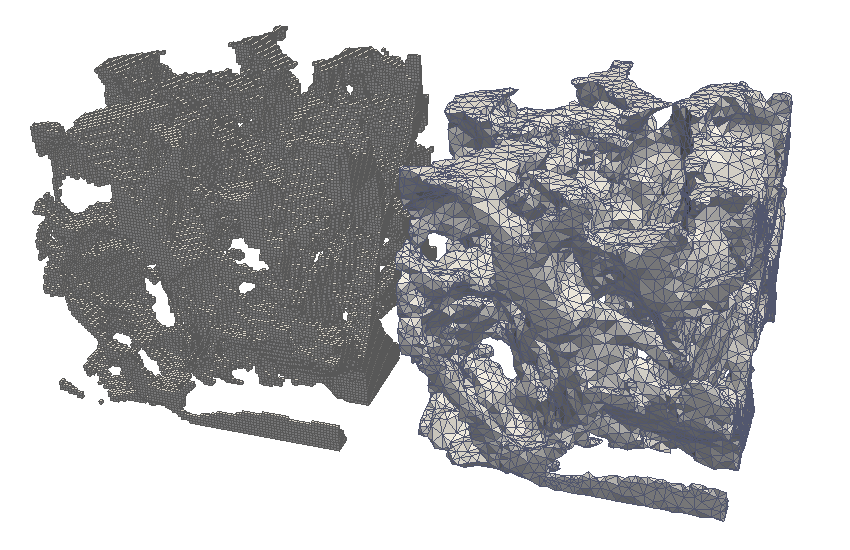}
\end{center}
\caption[]{Pore space corresponding to 0.45 mm linear size sandstone rock sample at saturation $s=0$. Left: $100^3$ voxel segmentation of the original CT scan. Right: tetrahedralization of the voxelized volume on the left.}
\label{fig:case_B}
\end{figure}

The data in Case A (figure~\ref{fig:case_A}) consists of $90^3$ voxels, whereas in case B (figure~\ref{fig:case_B}) the image consists of $100^3$ voxels.
Such small samples may be not representative for evaluating macroscopic flow properties of the original rock from which they were extracted.  
Since the main objective of the present study is comparison of different methods for the evaluation of relative permeability curves, the input data were chosen to run multiple numerical simulations on a desktop-size workstation in a reasonable time.  


For each saturation, the numerical solutions to the Stokes flow and Laplace equations can be obtained, for example, by means of a finite difference (FD) or a finite elements (FEM) scheme. 
The numerical evaluation of the surface integral in eq.~\eqref{eq:Lambda} for a voxelized geometry presents, however, a problem, which can be easily seen by taking the limiting case of a cylindrical pore space of radius $R$ and length $L$. 
In this case, the exact value of the surface area integral in eq.~\eqref{eq:Lambda}	is $2 \pi RL$, but its numerical value for a voxel geometry equals $4RL$, regardless of the voxel resolution.

For this reason, in order  to calculate the water-tight triangulated surface that includes the pore space in the voxelated image, we meshed the isosurface defining the pore-rock and fluid-fluid interface interfaces~\cite{Boissonnat05,cgal}. A new mesh is needed for each individual value of the saturation, both for drainage and imbibition.
Next, we generated a tetrahedralization of the volume contained in the closed surface.
Care was taken of removing isolated regions not connected to the inlet and outlet faces of the cube, and isolated surfaces not belonging to any of the tetrahedra.
The obtained  tetrahedra were then used to create a finite element mesh for the Comsol Multiphysics~\cite[]{COMSOL} Stokes and Laplace solvers.
A comparison between the voxelated images and their tetrahedralization is presented in figures~\ref{fig:case_A} and~\ref{fig:case_B}.
The tetrahedralization of the pore space preserves the original value of porosity to a high degree of accuracy, and allows for an accurate evaluation of the surface integral in eq.~\eqref{eq:Lambda}.   
As for the evaluation of the flow and integrals by means of the finite difference method, both Laplace and Stokes equations were solved on a mesh composed of the image pore voxels. 
The surface integral was evaluated either using the rectangular mesh, or with a marching-cubes algorithm~\cite[]{Lorensen:1987,MarchingCubes33:1995}.

\section{Numerical evaluation of the relative permeability curves}

We evaluate the relative permeability curves by means of five different numerical schemes, hereafter referred to as $m_i$,  $i=1,2,\ldots,5$, as detailed in Table~\ref{table:summary}.
\begin{table}
  \centering
  \begin{tabular}{|l|l|l|l|l|}
  \hline
  $ $ & Problem & Numerical scheme & Mesh type & Surface integration\\
  \hline
  $m_1$ & Stokes    & FEM & tetrahedra & n.a.\\
  $m_2$ & Potential & FEM & tetrahedra & quadratic \\
  $m_3$ & Stokes    & FD  & voxels     & n.a.\\
  $m_4$ & Potential & FD  & voxels     & staircase\\
  $m_5$ & Potential & FD  & voxels     & marching cubes\\
  \hline
\end{tabular}
\caption{Summary of the numerical schemes used for the evaluation of the relative permeabilities for case A and B (see Figures~\ref{fig:case_A} and~\ref{fig:case_B}).}\label{table:summary}
\end{table}
%
In FD scheme ($m_4$), the surface integral was evaluated on the interfaces between pore and solid voxels, whereas in the scheme ($m_5$) the integral was evaluated using a marching-cubes approximation of the surface between pores and solid.

\begin{figure}[h]
\begin{center}
\includegraphics[width=0.45\textwidth]{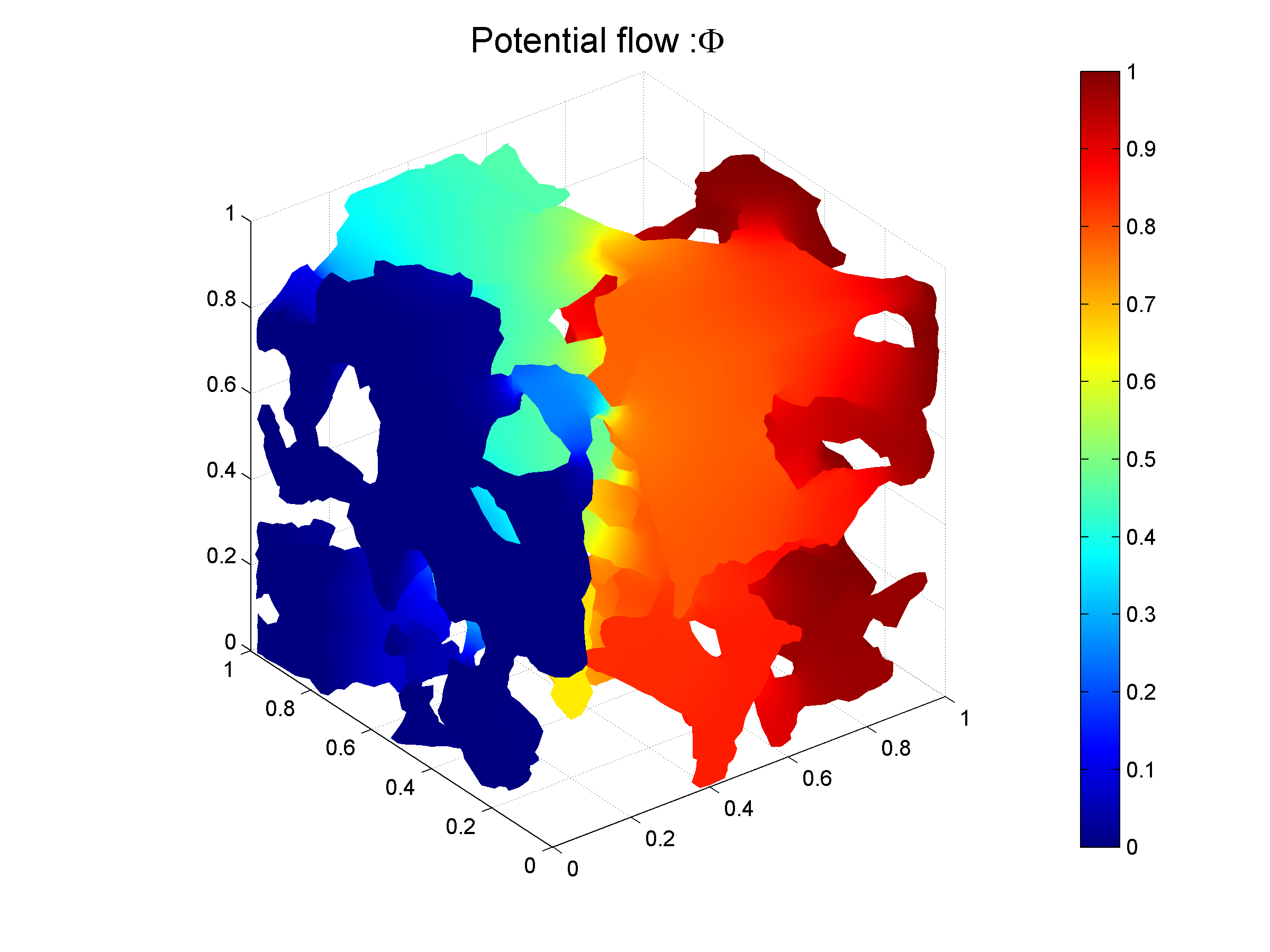}
\includegraphics[width=0.45\textwidth]{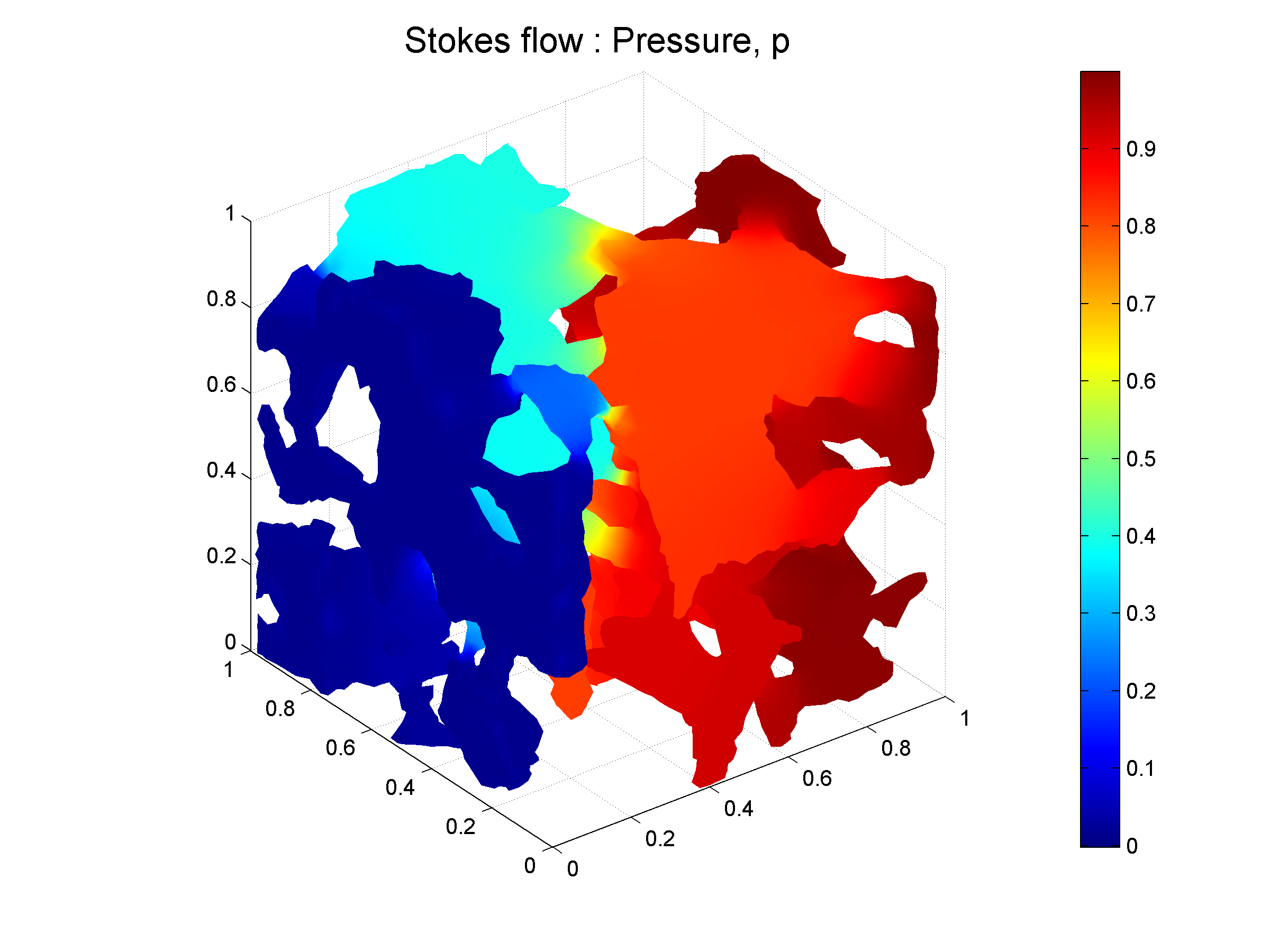}\\
\includegraphics[width=0.45\textwidth]{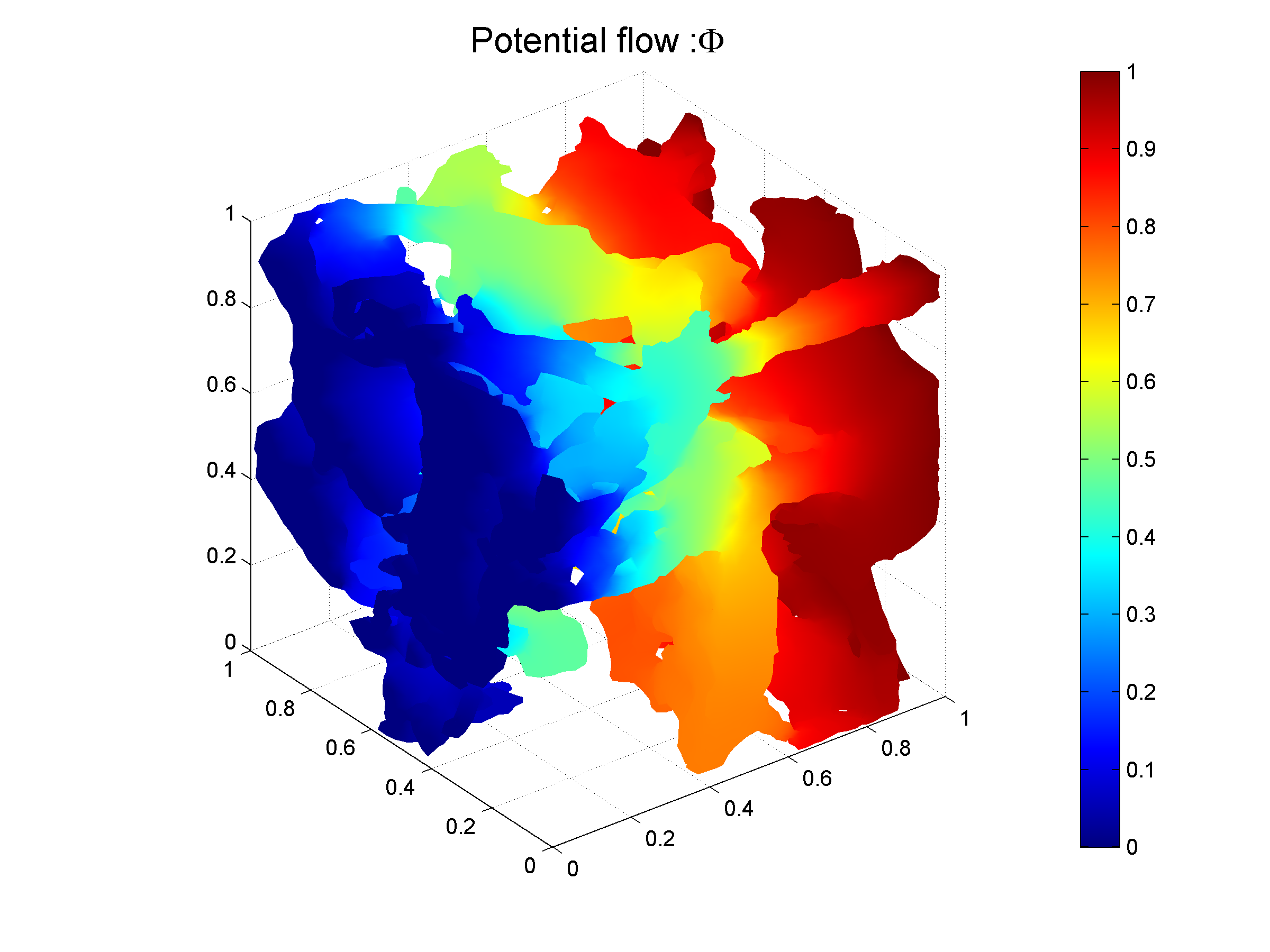}
\includegraphics[width=0.45\textwidth]{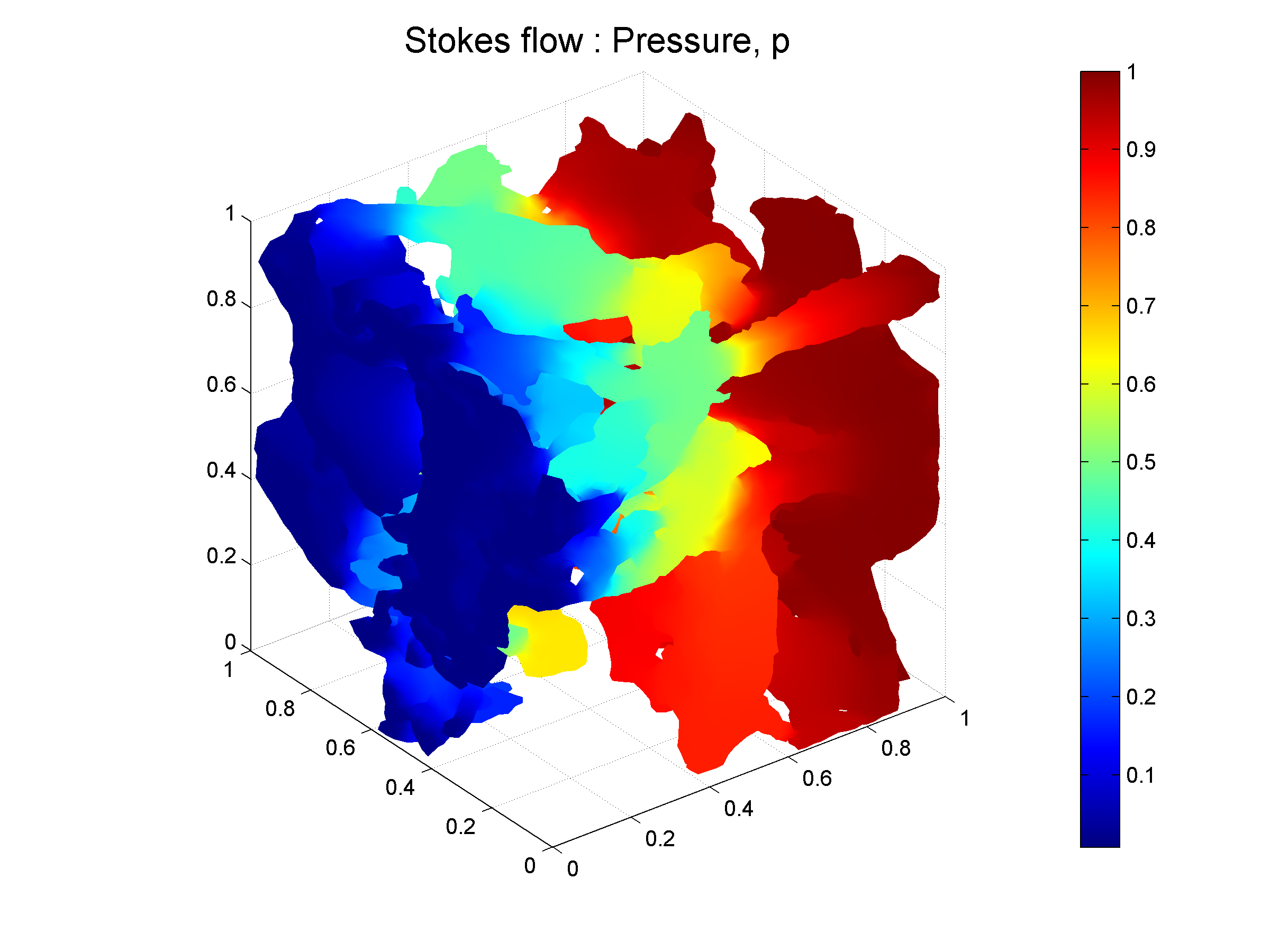}\\
\end{center}
\caption[]{Potential flow (left) and Stokes flow (right) FEM simulations in the $x$ direction for Case A (top) and case B (bottom).}
\label{fig:simulations}
\end{figure}

The boundary conditions for the potential flow ($m_2$, $m_4$, and $m_5$) were selected to have a unit potential drop across the opposite faces of the sample. For the Stokes flow problem ($m_1$, $m_3$), we set $p=0$ for the pressure on the opposite faces of the sample and added a unit constant body force along the direction of flow. Due to the linearity of the porblem, both the boundary pressure drop and body force approach are equivalent to each other.
For all numerical schemes, the flow was evaluated in the three orthogonal directions, $x$, $y$, and $z$.
Figure~\ref{fig:simulations} shows the pressure and potential fields evaluated by FEM numerical solutions ($m_1$, and $m_2$) for flow in the $x$ direction for both the Stokes and potential problems.  Note that visually the potential and pressure distributions are almost indistinguishable from each other.

The  permeability  for the Stokes flow was obtained by averaging the velocity vector components in the direction of the flow. The permeability for the potential flow case was obtained applying eq.~\eqref{eq:JKDPerm} to the potential flow solution.

\begin{figure}[h]
\begin{center}
\includegraphics[width=0.85\textwidth]{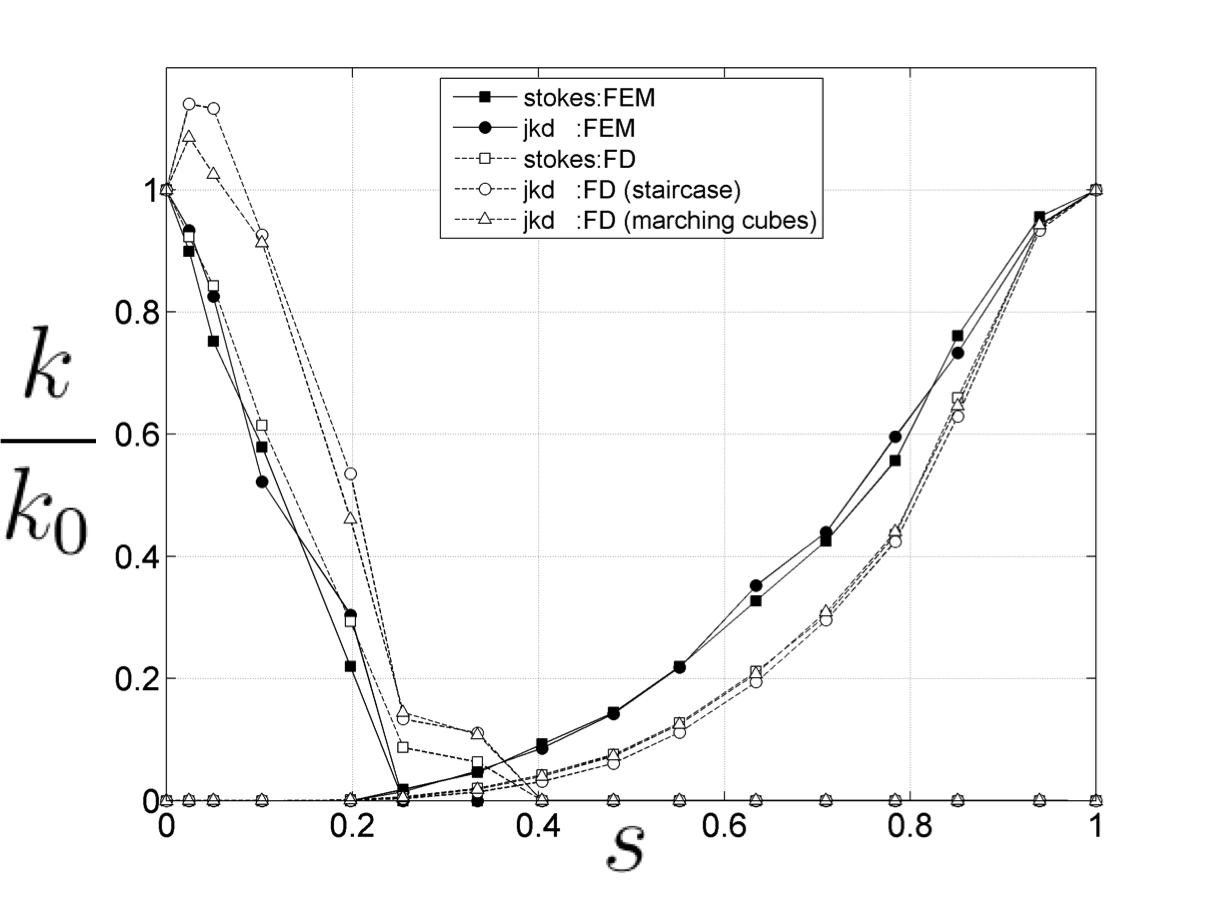}\\
\includegraphics[width=0.85\textwidth]{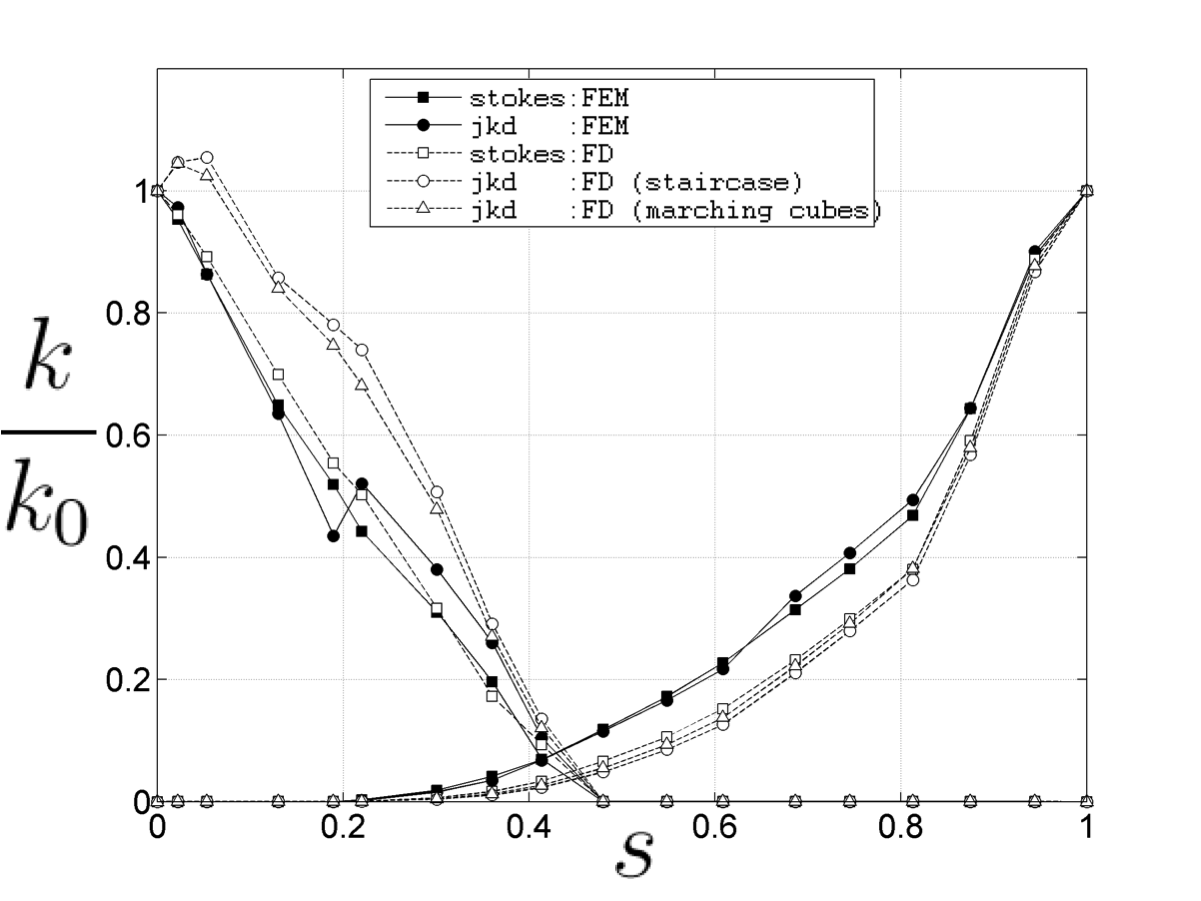}
\end{center}
\caption[]{Numerical evaluation of the relative permeability curves for Stokes flow and the Johnson-Koplik-Dashen estimates for case A and case B.}
\label{fig:relperm}
\end{figure}

Figure~\eqref{fig:relperm} shows the numerical evaluation of mean value of the relative permeability tensor diagonal elements for all numerical schemes.
As it can be seen, the JKD approximation in eq.~\eqref{eq:JKDPerm} ($m_2$, $m_4$, and $m_5$) is generally in good agreement with the Stokes flow calculation ($m_1$, $m_3$) for both case A and case B. The FD estimates of the JKD relative permeability for the   non-wetting phase ($m_4$, $m_5$), however, differ significantly for small saturations.  Moreover, both $m_4$ and $m_5$  relative permeability estimates  exceed unity for small wetting fluid saturation values, which is not physically meaningful.  
This outcome can be attributed to large variations of the effective $M$ values in FD simulations. 
Such nonphysical overshoot behavior has not been observed in the FEM simulations.  At the same time, for data set B, the FEM calculations of the JKD nonwetting relative permeability curve ($m_2$) displays a kink for $s \sim 0.18$, which can be attributed to some yet unresolved issues in the generation of the tetrahedral mesh.
Tables~\ref{tab:alpha_lambda_case_A} and \ref{tab:alpha_lambda_case_B} summarize the values of the tortuosity $\alpha_{\infty}$ and viscous length $\Lambda$ for the FEM computations. It can be observed that the tortuosity displays general monotonic changes as a function of saturation, whereas $\Lambda$ does not display such a general a monotonic behavior. A decreasing trend of $\alpha_{\infty}$ with pore volume $\phi$, as indicated by the computations, is expected: the precise structure of the spatial connectivity, however, is another important factor affecting the value of $\alpha_{\infty}$ and very little can be said a-priori about changes in connectivity as a function of change in saturation. 
Connectivity also strongly 
influences the characteristic viscous length value, which is roughly a measure of the (square root) surface area of the pore space at any given saturation. Also here the details of the pore geometry are extremely important and their effect on $\Lambda$ cannot be inferred in any simple way by geometric consideration alone.
Nonetheless, the constraint in eq.~\eqref{eq:condition} remains satisfied. 

\def\imagetop#1{\vtop{\null\hbox{#1}}}
\begin{table}[t]
\centering
\begin{tabular}{c}
\hline
case A  \\
\hline
\vspace{-3mm}\\
\vspace{-2mm}$x$-direction  \\
\imagetop{\begin{tiny}\begin{tabular}{|p{1cm}|p{1cm}|p{1cm}|p{1cm}|p{1cm}|p{1cm}|p{1cm}|p{1cm}|p{1cm}|p{1cm}|p{1cm}|p{1cm}|p{1cm}|}
\hline
\textbf{$s$} & \textbf{$\phi^{oil}$} & \textbf{$\phi^{wtr}$} & \textbf{$\alpha^{oil}$} & \textbf{$\alpha^{wtr}$} & \textbf{$\Lambda^{oil}$} & \textbf{$\Lambda^{wtr}$} & \textbf{$k^{oil}_{stokes}\times 10^{-6}$} & \textbf{$k^{wtr}_{stokes}\times 10^{-6}$} & \textbf{$k^{oil}_{JKD}\times 10^{-6}$} & \textbf{$k^{wtr}_{JKD}\times 10^{-6}$} & \textbf{$M^{oil}$} & \textbf{$M^{wtr}$}\\\hline
0.0000 & 0.2862 & 0.1398 & 3.2215 & 0.0000 & 0.0380 & 0.0000 & 22.0961 & 0.0000 & 16.0388 & 0.0000 & 1.3777 & NaN\\\hline
0.0246 & 0.2750 & 0.1398 & 3.9474 & 0.0000 & 0.0429 & 0.0000 & 19.7175 & 0.0000 & 15.9952 & 0.0000 & 1.2327 & NaN\\\hline
0.0505 & 0.2608 & 0.1398 & 4.4033 & 0.0000 & 0.0442 & 0.0000 & 16.2550 & 0.0000 & 14.4922 & 0.0000 & 1.1216 & NaN\\\hline
0.1031 & 0.2327 & 0.1398 & 5.5683 & 0.0000 & 0.0398 & 0.0000 & 12.1761 & 0.0000 & 8.2876 & 0.0000 & 1.4692 & NaN\\\hline
0.1981 & 0.1398 & 0.1398 & 7.8101 & 0.0000 & 0.0499 & 0.0000 & 4.6498 & 0.0000 & 5.5721 & 0.0000 & 0.8345 & NaN\\\hline
0.2548 & 0.1398 & 0.0803 & 0.0000 & 13.4248 & 0.0000 & 0.0132 & 0.0000 & 0.2696 & 0.0000 & 0.1293 & NaN & 2.0839\\\hline
0.3347 & 0.1398 & 0.1145 & 0.0000 & 6.4513 & 0.0000 & 0.0207 & 0.0000 & 1.1199 & 0.0000 & 0.9546 & NaN & 1.1731\\\hline
0.4039 & 0.1398 & 0.1405 & 0.0000 & 4.6189 & 0.0000 & 0.0200 & 0.0000 & 2.2169 & 0.0000 & 1.5142 & NaN & 1.4640\\\hline
0.4807 & 0.1398 & 0.1624 & 0.0000 & 3.9344 & 0.0000 & 0.0241 & 0.0000 & 3.6047 & 0.0000 & 3.0021 & NaN & 1.2007\\\hline
0.5520 & 0.1398 & 0.1824 & 0.0000 & 3.5504 & 0.0000 & 0.0265 & 0.0000 & 5.3575 & 0.0000 & 4.4951 & NaN & 1.1919\\\hline
0.6344 & 0.1398 & 0.2043 & 0.0000 & 3.2310 & 0.0000 & 0.0302 & 0.0000 & 8.0858 & 0.0000 & 7.1999 & NaN & 1.1230\\\hline
0.7097 & 0.1398 & 0.2043 & 0.0000 & 0.0000 & 0.0000 & 0.0000 & 0.0000 & 0.0000 & 0.0000 & 0.0000 & NaN & NaN\\\hline
0.7842 & 0.1398 & 0.2391 & 0.0000 & 3.0977 & 0.0000 & 0.0347 & 0.0000 & 13.7151 & 0.0000 & 11.6440 & NaN & 1.1779\\\hline
0.8513 & 0.1398 & 0.2536 & 0.0000 & 3.0696 & 0.0000 & 0.0373 & 0.0000 & 19.9447 & 0.0000 & 14.3685 & NaN & 1.3881\\\hline
0.9393 & 0.1398 & 0.2733 & 0.0000 & 3.1581 & 0.0000 & 0.0383 & 0.0000 & 21.5681 & 0.0000 & 15.8340 & NaN & 1.3621\\\hline
1.0000 & 0.1398 & 0.2862 & 0.0000 & 3.2215 & 0.0000 & 0.0380 & 0.0000 & 22.0961 & 0.0000 & 16.0388 & NaN & 1.3777\\\hline
\end{tabular}
\end{tiny}
} \\
\vspace{-3mm}\\
\vspace{-2mm}$y$-direction  \\
\imagetop{\begin{tiny}\begin{tabular}{|p{1cm}|p{1cm}|p{1cm}|p{1cm}|p{1cm}|p{1cm}|p{1cm}|p{1cm}|p{1cm}|p{1cm}|p{1cm}|p{1cm}|p{1cm}|}
\hline
\textbf{$s$} & \textbf{$\phi^{oil}$} & \textbf{$\phi^{wtr}$} & \textbf{$\alpha^{oil}$} & \textbf{$\alpha^{wtr}$} & \textbf{$\Lambda^{oil}$} & \textbf{$\Lambda^{wtr}$} & \textbf{$k^{oil}_{stokes}\times 10^{-6}$} & \textbf{$k^{wtr}_{stokes}\times 10^{-6}$} & \textbf{$k^{oil}_{JKD}\times 10^{-6}$} & \textbf{$k^{wtr}_{JKD}\times 10^{-6}$} & \textbf{$M^{oil}$} & \textbf{$M^{wtr}$}\\\hline
0.0000 & 0.2862 & 0.1398 & 2.2008 & 0.0000 & 0.0418 & 0.0000 & 32.9858 & 0.0000 & 28.4339 & 0.0000 & 1.1601 & NaN\\\hline
0.0246 & 0.2750 & 0.1398 & 2.4311 & 0.0000 & 0.0425 & 0.0000 & 29.8501 & 0.0000 & 25.5725 & 0.0000 & 1.1673 & NaN\\\hline
0.0505 & 0.2608 & 0.1398 & 2.6760 & 0.0000 & 0.0424 & 0.0000 & 24.6998 & 0.0000 & 21.9435 & 0.0000 & 1.1256 & NaN\\\hline
0.1031 & 0.2327 & 0.1398 & 3.2439 & 0.0000 & 0.0397 & 0.0000 & 19.3292 & 0.0000 & 14.1141 & 0.0000 & 1.3695 & NaN\\\hline
0.1981 & 0.1398 & 0.1398 & 9.2754 & 0.0000 & 0.0516 & 0.0000 & 4.6659 & 0.0000 & 5.0148 & 0.0000 & 0.9304 & NaN\\\hline
0.2548 & 0.1398 & 0.0803 & 0.0000 & 6.6070 & 0.0000 & 0.0156 & 0.0000 & 0.6650 & 0.0000 & 0.3715 & NaN & 1.7900\\\hline
0.3347 & 0.1398 & 0.1145 & 0.0000 & 5.7809 & 0.0000 & 0.0222 & 0.0000 & 1.4744 & 0.0000 & 1.2236 & NaN & 1.2050\\\hline
0.4039 & 0.1398 & 0.1405 & 0.0000 & 3.3827 & 0.0000 & 0.0230 & 0.0000 & 3.4940 & 0.0000 & 2.7462 & NaN & 1.2723\\\hline
0.4807 & 0.1398 & 0.1624 & 0.0000 & 3.0771 & 0.0000 & 0.0250 & 0.0000 & 5.2258 & 0.0000 & 4.1136 & NaN & 1.2704\\\hline
0.5520 & 0.1398 & 0.1824 & 0.0000 & 2.6877 & 0.0000 & 0.0271 & 0.0000 & 7.9025 & 0.0000 & 6.2448 & NaN & 1.2655\\\hline
0.6344 & 0.1398 & 0.2043 & 0.0000 & 2.4884 & 0.0000 & 0.0302 & 0.0000 & 11.0166 & 0.0000 & 9.3833 & NaN & 1.1741\\\hline
0.7097 & 0.1398 & 0.2230 & 0.0000 & 2.4388 & 0.0000 & 0.0322 & 0.0000 & 12.9345 & 0.0000 & 11.8313 & NaN & 1.0932\\\hline
0.7842 & 0.1398 & 0.2391 & 0.0000 & 2.3506 & 0.0000 & 0.0340 & 0.0000 & 15.9074 & 0.0000 & 14.7255 & NaN & 1.0803\\\hline
0.8513 & 0.1398 & 0.2536 & 0.0000 & 2.3050 & 0.0000 & 0.0363 & 0.0000 & 21.5552 & 0.0000 & 18.1037 & NaN & 1.1907\\\hline
0.9393 & 0.1398 & 0.2733 & 0.0000 & 2.1982 & 0.0000 & 0.0406 & 0.0000 & 30.0686 & 0.0000 & 25.6279 & NaN & 1.1733\\\hline
1.0000 & 0.1398 & 0.2862 & 0.0000 & 2.2008 & 0.0000 & 0.0418 & 0.0000 & 32.9858 & 0.0000 & 28.4339 & NaN & 1.1601\\\hline
\end{tabular}
\end{tiny}} \\
\vspace{-3mm}\\
\vspace{-2mm}$z$-direction  \\
\imagetop{\begin{tiny}\begin{tabular}{|p{1cm}|p{1cm}|p{1cm}|p{1cm}|p{1cm}|p{1cm}|p{1cm}|p{1cm}|p{1cm}|p{1cm}|p{1cm}|p{1cm}|p{1cm}|}
\hline
\textbf{$s$} & \textbf{$\phi^{oil}$} & \textbf{$\phi^{wtr}$} & \textbf{$\alpha^{oil}$} & \textbf{$\alpha^{wtr}$} & \textbf{$\Lambda^{oil}$} & \textbf{$\Lambda^{wtr}$} & \textbf{$k^{oil}_{stokes}\times 10^{-6}$} & \textbf{$k^{wtr}_{stokes}\times 10^{-6}$} & \textbf{$k^{oil}_{JKD}\times 10^{-6}$} & \textbf{$k^{wtr}_{JKD}\times 10^{-6}$} & \textbf{$M^{oil}$} & \textbf{$M^{wtr}$}\\\hline
0.0000 & 0.2860 & 0.1398 & 2.3506 & 0.0000 & 0.0420 & 0.0000 & 32.9003 & 0.0000 & 26.9091 & 0.0000 & 1.2235 & NaN\\\hline
0.0246 & 0.2750 & 0.1398 & 2.7516 & 0.0000 & 0.0448 & 0.0000 & 29.5074 & 0.0000 & 25.0861 & 0.0000 & 1.1762 & NaN\\\hline
0.0505 & 0.2608 & 0.1398 & 3.1635 & 0.0000 & 0.0468 & 0.0000 & 25.1819 & 0.0000 & 22.5282 & 0.0000 & 1.1178 & NaN\\\hline
0.1031 & 0.2327 & 0.1398 & 3.6185 & 0.0000 & 0.0430 & 0.0000 & 19.4750 & 0.0000 & 14.8542 & 0.0000 & 1.3111 & NaN\\\hline
0.1981 & 0.1398 & 0.1398 & 4.0768 & 0.0000 & 0.0509 & 0.0000 & 10.0152 & 0.0000 & 11.1119 & 0.0000 & 0.9013 & NaN\\\hline
0.2548 & 0.1398 & 0.0803 & 0.0000 & 6.1335 & 0.0000 & 0.0181 & 0.0000 & 0.6766 & 0.0000 & 0.5389 & NaN & 1.2556\\\hline
0.3347 & 0.1398 & 0.1145 & 0.0000 & 5.0763 & 0.0000 & 0.0217 & 0.0000 & 1.4611 & 0.0000 & 1.3215 & NaN & 1.1056\\\hline
0.4039 & 0.1398 & 0.1405 & 0.0000 & 4.4070 & 0.0000 & 0.0215 & 0.0000 & 2.4553 & 0.0000 & 1.8495 & NaN & 1.3275\\\hline
0.4807 & 0.1398 & 0.1624 & 0.0000 & 3.7336 & 0.0000 & 0.0235 & 0.0000 & 3.8285 & 0.0000 & 3.0037 & NaN & 1.2746\\\hline
0.5520 & 0.1398 & 0.1824 & 0.0000 & 3.2120 & 0.0000 & 0.0260 & 0.0000 & 6.0558 & 0.0000 & 4.8097 & NaN & 1.2591\\\hline
0.6344 & 0.1398 & 0.2046 & 0.0000 & 2.7093 & 0.0000 & 0.0301 & 0.0000 & 9.6236 & 0.0000 & 8.5518 & NaN & 1.1234\\\hline
0.7097 & 0.1398 & 0.2230 & 0.0000 & 2.5212 & 0.0000 & 0.0325 & 0.0000 & 13.3291 & 0.0000 & 11.6893 & NaN & 1.1403\\\hline
0.7842 & 0.1398 & 0.2391 & 0.0000 & 2.3934 & 0.0000 & 0.0360 & 0.0000 & 19.3948 & 0.0000 & 16.2098 & NaN & 1.1965\\\hline
0.8513 & 0.1398 & 0.2536 & 0.0000 & 2.3462 & 0.0000 & 0.0383 & 0.0000 & 25.4538 & 0.0000 & 19.8345 & NaN & 1.2833\\\hline
0.9393 & 0.1398 & 0.2733 & 0.0000 & 2.2752 & 0.0000 & 0.0414 & 0.0000 & 32.4782 & 0.0000 & 25.7770 & NaN & 1.2600\\\hline
1.0000 & 0.1398 & 0.2860 & 0.0000 & 2.3506 & 0.0000 & 0.0420 & 0.0000 & 32.9003 & 0.0000 & 26.9091 & NaN & 1.2235\\\hline
\end{tabular}
\end{tiny}} \\
\end{tabular}

\caption{\footnotesize Values of porosity $\phi$, tortuosity $\alpha_{\infty}$, viscous length $\Lambda$, Stokes relative permeability $k$ (FEM method), and $M$ as a function of saturation, $s$, for draining and imbibition for the case~A.}
\label{tab:alpha_lambda_case_A}
\end{table}

\begin{table}[t]
\vspace{-5mm}
\centering
\begin{tabular}{c}
\hline
case B  \\
\hline
\vspace{-5mm}\\
\vspace{-3mm}$x$-direction  \\
\imagetop{\begin{tiny}\begin{tabular}{|p{1cm}|p{1cm}|p{1cm}|p{1cm}|p{1cm}|p{1cm}|p{1cm}|p{1cm}|p{1cm}|p{1cm}|p{1cm}|p{1cm}|p{1cm}|}
\hline
{$s$} & {$\phi^{oil}$} & {$\phi^{wtr}$} & {$\alpha^{oil}$} & {$\alpha^{wtr}$} & {$\Lambda^{oil}$} & {$\Lambda^{wtr}$} & {$k^{oil}_{stokes}\times 10^{-6}$} & {$k^{wtr}_{stokes}\times 10^{-6}$} & {$k^{oil}_{JKD}\times 10^{-6}$} & {$k^{wtr}_{JKD}\times 10^{-6}$} & {$M^{oil}$} & {$M^{wtr}$}\\\hline
0.0000 & 0.3283 & 0.0788 & 2.1274 & 0.0000 & 0.0417 & 0.0000 & 45.3083 & 0.0000 & 33.6124 & 0.0000 & 1.3480 & NaN\\\hline
0.0222 & 0.3184 & 0.0788 & 2.2342 & 0.0000 & 0.0429 & 0.0000 & 43.2532 & 0.0000 & 32.7742 & 0.0000 & 1.3197 & NaN\\\hline
0.0535 & 0.3047 & 0.0788 & 2.5633 & 0.0000 & 0.0446 & 0.0000 & 38.1059 & 0.0000 & 29.6166 & 0.0000 & 1.2866 & NaN\\\hline
0.1301 & 0.2732 & 0.0788 & 3.2086 & 0.0000 & 0.0458 & 0.0000 & 29.7895 & 0.0000 & 22.3242 & 0.0000 & 1.3344 & NaN\\\hline
0.1896 & 0.2521 & 0.0788 & 4.2438 & 0.0000 & 0.0353 & 0.0000 & 24.1109 & 0.0000 & 9.1773 & 0.0000 & 2.6073 & NaN\\\hline
0.2204 & 0.2364 & 0.0527 & 4.6174 & 12.5101 & 0.0582 & 0.0155 & 22.2675 & 0.2625 & 21.6762 & 0.1272 & 1.0276 & 2.0657\\\hline
0.3009 & 0.1873 & 0.1047 & 4.8449 & 6.9540 & 0.0605 & 0.0191 & 17.6274 & 0.9549 & 17.7168 & 0.6873 & 0.9950 & 1.3895\\\hline
0.3607 & 0.1636 & 0.1361 & 5.0389 & 4.7420 & 0.0602 & 0.0197 & 13.4267 & 2.1335 & 14.7230 & 1.3980 & 0.9120 & 1.5261\\\hline
0.4138 & 0.0788 & 0.1545 & 2.9422 & 3.9074 & 0.0592 & 0.0229 & 9.7478 & 3.2188 & 11.7356 & 2.5844 & 0.8306 & 1.2455\\\hline
0.4796 & 0.0788 & 0.1824 & 0.0000 & 2.9846 & 0.0000 & 0.0244 & 0.0000 & 5.6659 & 0.0000 & 4.5388 & NaN & 1.2483\\\hline
0.5487 & 0.0788 & 0.2068 & 0.0000 & 2.7446 & 0.0000 & 0.0263 & 0.0000 & 7.7567 & 0.0000 & 6.5247 & NaN & 1.1888\\\hline
0.6090 & 0.0788 & 0.2068 & 0.0000 & 0.0000 & 0.0000 & 0.0000 & 0.0000 & 0.0000 & 0.0000 & 0.0000 & NaN & NaN\\\hline
0.6867 & 0.0788 & 0.2482 & 0.0000 & 2.4690 & 0.0000 & 0.0298 & 0.0000 & 12.8325 & 0.0000 & 11.1420 & NaN & 1.1517\\\hline
0.7456 & 0.0788 & 0.2640 & 0.0000 & 2.3973 & 0.0000 & 0.0313 & 0.0000 & 15.3481 & 0.0000 & 13.4547 & NaN & 1.1407\\\hline
0.8131 & 0.0788 & 0.2815 & 0.0000 & 2.3210 & 0.0000 & 0.0334 & 0.0000 & 19.0626 & 0.0000 & 16.9418 & NaN & 1.1252\\\hline
0.8752 & 0.0788 & 0.2977 & 0.0000 & 2.1580 & 0.0000 & 0.0376 & 0.0000 & 32.2647 & 0.0000 & 24.4282 & NaN & 1.3208\\\hline
0.9444 & 0.0788 & 0.3152 & 0.0000 & 2.0953 & 0.0000 & 0.0415 & 0.0000 & 43.9242 & 0.0000 & 32.4062 & NaN & 1.3554\\\hline
1.0000 & 0.0788 & 0.3283 & 0.0000 & 2.1274 & 0.0000 & 0.0417 & 0.0000 & 45.3083 & 0.0000 & 33.6124 & NaN & 1.3480\\\hline
\end{tabular}
\end{tiny}} \\
\vspace{-5mm}\\
\vspace{-3mm}$y$-direction  \\
\imagetop{\begin{tiny}\begin{tabular}{|p{1cm}|p{1cm}|p{1cm}|p{1cm}|p{1cm}|p{1cm}|p{1cm}|p{1cm}|p{1cm}|p{1cm}|p{1cm}|p{1cm}|p{1cm}|}
\hline
{$s$} & {$\phi^{oil}$} & {$\phi^{wtr}$} & {$\alpha^{oil}$} & {$\alpha^{wtr}$} & {$\Lambda^{oil}$} & {$\Lambda^{wtr}$} & {$k^{oil}_{stokes}\times 10^{-6}$} & {$k^{wtr}_{stokes}\times 10^{-6}$} & {$k^{oil}_{JKD}\times 10^{-6}$} & {$k^{wtr}_{JKD}\times 10^{-6}$} & {$M^{oil}$} & {$M^{wtr}$}\\\hline
0.0000 & 0.3283 & 0.1636 & 1.8537 & 0.0000 & 0.0431 & 0.0000 & 46.2177 & 0.0000 & 41.0462 & 0.0000 & 1.1260 & NaN\\\hline
0.0222 & 0.3184 & 0.1636 & 1.9184 & 0.0000 & 0.0440 & 0.0000 & 43.8856 & 0.0000 & 40.1172 & 0.0000 & 1.0939 & NaN\\\hline
0.0535 & 0.3047 & 0.1636 & 2.0149 & 0.0000 & 0.0429 & 0.0000 & 39.7709 & 0.0000 & 34.7468 & 0.0000 & 1.1446 & NaN\\\hline
0.1301 & 0.2732 & 0.1636 & 2.7170 & 0.0000 & 0.0426 & 0.0000 & 28.1039 & 0.0000 & 22.8202 & 0.0000 & 1.2315 & NaN\\\hline
0.1896 & 0.2521 & 0.1636 & 3.6725 & 0.0000 & 0.0504 & 0.0000 & 20.7261 & 0.0000 & 21.6665 & 0.0000 & 0.9494 & NaN\\\hline
0.2204 & 0.2364 & 0.0527 & 3.9164 & 20.8180 & 0.0479 & 0.0170 & 16.9475 & 0.1643 & 17.3365 & 0.0916 & 0.9779 & 1.7939\\\hline
0.3009 & 0.1873 & 0.1047 & 5.6321 & 6.5054 & 0.0475 & 0.0170 & 10.3014 & 0.8529 & 9.3874 & 0.5783 & 1.0974 & 1.4749\\\hline
0.3607 & 0.1636 & 0.1361 & 9.2010 & 4.9712 & 0.0454 & 0.0187 & 5.0568 & 1.6367 & 4.5861 & 1.2018 & 1.1026 & 1.3619\\\hline
0.4138 & 0.1636 & 0.1545 & 0.0000 & 4.0250 & 0.0000 & 0.0229 & 0.0000 & 2.9487 & 0.0000 & 2.5262 & NaN & 1.1672\\\hline
0.4796 & 0.1636 & 0.1824 & 0.0000 & 3.1654 & 0.0000 & 0.0234 & 0.0000 & 5.0860 & 0.0000 & 3.9385 & NaN & 1.2914\\\hline
0.5487 & 0.1636 & 0.2068 & 0.0000 & 2.7127 & 0.0000 & 0.0244 & 0.0000 & 7.7957 & 0.0000 & 5.6729 & NaN & 1.3742\\\hline
0.6090 & 0.1636 & 0.2254 & 0.0000 & 2.4668 & 0.0000 & 0.0265 & 0.0000 & 10.6008 & 0.0000 & 8.0289 & NaN & 1.3203\\\hline
0.6867 & 0.1636 & 0.2482 & 0.0000 & 2.2194 & 0.0000 & 0.0316 & 0.0000 & 15.7903 & 0.0000 & 13.9263 & NaN & 1.1339\\\hline
0.7456 & 0.1636 & 0.2640 & 0.0000 & 2.1266 & 0.0000 & 0.0341 & 0.0000 & 20.0870 & 0.0000 & 18.0202 & NaN & 1.1147\\\hline
0.8131 & 0.1636 & 0.2815 & 0.0000 & 2.0505 & 0.0000 & 0.0352 & 0.0000 & 24.0628 & 0.0000 & 21.2649 & NaN & 1.1316\\\hline
0.8752 & 0.1636 & 0.2977 & 0.0000 & 1.9823 & 0.0000 & 0.0376 & 0.0000 & 29.8248 & 0.0000 & 26.5431 & NaN & 1.1236\\\hline
0.9444 & 0.1636 & 0.3152 & 0.0000 & 1.8824 & 0.0000 & 0.0417 & 0.0000 & 39.1592 & 0.0000 & 36.4566 & NaN & 1.0741\\\hline
1.0000 & 0.1636 & 0.3283 & 0.0000 & 1.8537 & 0.0000 & 0.0431 & 0.0000 & 46.2177 & 0.0000 & 41.0462 & NaN & 1.1260\\\hline
\end{tabular}
\end{tiny}} \\
\vspace{-5mm}\\
\vspace{-3mm}$z$-direction  \\
\imagetop{\begin{tiny}\begin{tabular}{|p{1cm}|p{1cm}|p{1cm}|p{1cm}|p{1cm}|p{1cm}|p{1cm}|p{1cm}|p{1cm}|p{1cm}|p{1cm}|p{1cm}|p{1cm}|}
\hline
{$s$} & {$\phi^{oil}$} & {$\phi^{wtr}$} & {$\alpha^{oil}$} & {$\alpha^{wtr}$} & {$\Lambda^{oil}$} & {$\Lambda^{wtr}$} & {$k^{oil}_{stokes}\times 10^{-6}$} & {$k^{wtr}_{stokes}\times 10^{-6}$} & {$k^{oil}_{JKD}\times 10^{-6}$} & {$k^{wtr}_{JKD}\times 10^{-6}$} & {$M^{oil}$} & {$M^{wtr}$}\\\hline
0.0000 & 0.3283 & 0.1636 & 2.0232 & 0.0000 & 0.0433 & 0.0000 & 48.7885 & 0.0000 & 38.0016 & 0.0000 & 1.2839 & NaN\\\hline
0.0222 & 0.3184 & 0.1636 & 2.0629 & 0.0000 & 0.0436 & 0.0000 & 46.5167 & 0.0000 & 36.7082 & 0.0000 & 1.2672 & NaN\\\hline
0.0535 & 0.3047 & 0.1636 & 2.2153 & 0.0000 & 0.0437 & 0.0000 & 43.2053 & 0.0000 & 32.8526 & 0.0000 & 1.3151 & NaN\\\hline
0.1301 & 0.2732 & 0.1636 & 2.4740 & 0.0000 & 0.0438 & 0.0000 & 33.1981 & 0.0000 & 26.4477 & 0.0000 & 1.2552 & NaN\\\hline
0.1896 & 0.2521 & 0.1636 & 2.8542 & 0.0000 & 0.0407 & 0.0000 & 26.4166 & 0.0000 & 18.1702 & 0.0000 & 1.4428 & NaN\\\hline
0.2204 & 0.2364 & 0.1636 & 3.4026 & 0.0000 & 0.0476 & 0.0000 & 22.9667 & 0.0000 & 19.6920 & 0.0000 & 1.1667 & NaN\\\hline
0.3009 & 0.1873 & 0.1047 & 4.3049 & 8.1797 & 0.0538 & 0.0182 & 15.6188 & 0.7678 & 15.7537 & 0.5301 & 0.9914 & 1.4483\\\hline
0.3607 & 0.1636 & 0.1361 & 6.1125 & 4.6589 & 0.0546 & 0.0191 & 9.0795 & 2.0762 & 9.9713 & 1.3296 & 0.9106 & 1.5615\\\hline
0.4138 & 0.1636 & 0.1545 & 0.0000 & 3.8298 & 0.0000 & 0.0224 & 0.0000 & 3.5063 & 0.0000 & 2.5315 & NaN & 1.3850\\\hline
0.4796 & 0.1636 & 0.1824 & 0.0000 & 3.1786 & 0.0000 & 0.0250 & 0.0000 & 5.8631 & 0.0000 & 4.4897 & NaN & 1.3059\\\hline
0.5487 & 0.1636 & 0.2068 & 0.0000 & 2.7678 & 0.0000 & 0.0265 & 0.0000 & 8.6007 & 0.0000 & 6.5572 & NaN & 1.3116\\\hline
0.6090 & 0.1636 & 0.2254 & 0.0000 & 2.5382 & 0.0000 & 0.0277 & 0.0000 & 11.5475 & 0.0000 & 8.5284 & NaN & 1.3540\\\hline
0.6867 & 0.1636 & 0.2482 & 0.0000 & 2.3799 & 0.0000 & 0.0315 & 0.0000 & 15.4805 & 0.0000 & 12.9350 & NaN & 1.1968\\\hline
0.7456 & 0.1636 & 0.2640 & 0.0000 & 2.3282 & 0.0000 & 0.0318 & 0.0000 & 17.9569 & 0.0000 & 14.3319 & NaN & 1.2529\\\hline
0.8131 & 0.1636 & 0.2815 & 0.0000 & 2.2504 & 0.0000 & 0.0334 & 0.0000 & 22.5232 & 0.0000 & 17.4200 & NaN & 1.2930\\\hline
0.8752 & 0.1636 & 0.2977 & 0.0000 & 2.1448 & 0.0000 & 0.0353 & 0.0000 & 28.1344 & 0.0000 & 21.6038 & NaN & 1.3023\\\hline
0.9444 & 0.1636 & 0.3152 & 0.0000 & 2.0298 & 0.0000 & 0.0410 & 0.0000 & 42.2662 & 0.0000 & 32.6128 & NaN & 1.2960\\\hline
1.0000 & 0.1636 & 0.3283 & 0.0000 & 2.0232 & 0.0000 & 0.0433 & 0.0000 & 48.7885 & 0.0000 & 38.0016 & NaN & 1.2839\\\hline
\end{tabular}
\end{tiny}} \\
\end{tabular}

\caption{\footnotesize Values of porosity $\phi$, tortuosity $\alpha_{\infty}$, viscous length $\Lambda$, Stokes relative permeability $k$  (FEM method), and $M$ as a function of saturation, $s$, for draining and imbibition for the case~B.}
\label{tab:alpha_lambda_case_B}
\end{table}

It is important to note that, as the voxelated and tetrahedralized images (see figures~\ref{fig:case_A} and~\ref{fig:case_B}) are not identical copies of each other. Thus, it would be unrealistic to expect a close agreement between the Stokes values of permeability for the FD (voxelated image) and FEM (tetrahedralized image) methods.

\begin{figure}[h]
\begin{center}
\includegraphics[width=12cm]{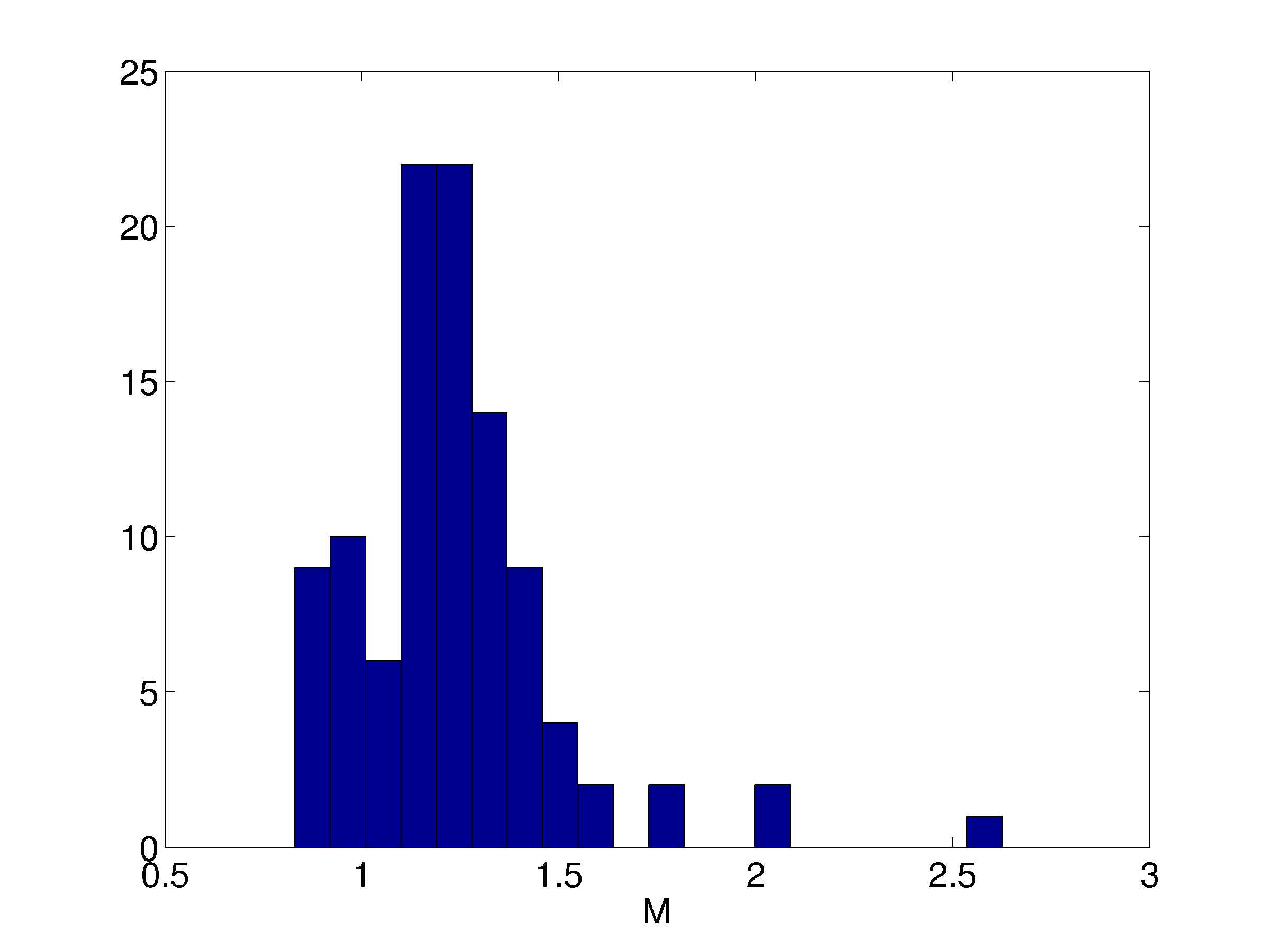}
\end{center}
\caption[]{Histogram of the M values for the teo cases A and B combined.}
\label{fig:M_hist}
\end{figure}

Finally, figure~\ref{fig:M_hist} shows a histogram of the $M$ values for the cases A and B combined. The mean value of $M=1.23$ is close to the theoretically predicted value $M \sim 1$ \cite[]{JKD87,BJ87,CKS88,CORBOOK}, and
the standard deviation is relatively small, $\sigma = 0.26$.

\section{Conclusions}
We have investigated the consequences of replacing solution of Stokes equations by a Laplace equation with respect to estimation of relative permeability curves.
The approximate relative permeability curves obtained by the JKD method for the two 3D microtomography images presented in this work is very close to the  Stokes curves.
The computational time needed to obtain the JKD solutions were, in this study, at least 20 times smaller than for solving the corresponding Stokes flow solutions.
Hence, expensive Stokes flow simulations can be replaced by calculating volume and surface averaged integrals of the squared modulus of a potential flow solution (see eq.~\eqref{eq:JKDPerm}). We also note here that the mesh generation does not represent a significant overburden on the overall computational cost.
The JKD estimate of relative permeability presented in this work has demonstrated a great potential for expedient evaluation of rock flow properties from micro tomography data.

\section*{Acknowledgements}\label{Sec6}%
This work has been performed at Lawrence Berkeley National Laboratory (LBNL) of
the U.S. Department of Energy (DOE) under Contract No. DE-AC02-05CH11231.
Funding for this project is provided by Research Partnership to Secure Energy for America (RPSEA) through the ``Ultra-Deepwater and Unconventional Natural Gas and Other Petroleum Resources'' program authorized by the U.S. Energy Policy Act of 2005. RPSEA (www.rpsea.org) is a nonprofit corporation whose mission is to provide a stewardship role in ensuring the focused research, development and deployment of safe and environmentally responsible technology that can effectively deliver hydrocarbons from domestic resources to the citizens of the United States. RPSEA, operating as a consortium of premier U.S. energy research universities, industry, and independent research organizations, manages the program under a contract with the U.S. DOE's National Energy Technology Laboratory.   The micro tomography data used in this study were obtained by J. Ajo-Franklin of LBNL at the Advanced Light Source facility of the Lawrence Berkeley National Laboratory.  The authors are thankful to Dr. Stefan Finsterle of LBNL for critical review of the manuscript.

\newpage
\newpage

\bibliographystyle{amsplain}

\providecommand{\bysame}{\leavevmode\hbox to3em{\hrulefill}\thinspace}
\providecommand{\MR}{\relax\ifhmode\unskip\space\fi MR }
\providecommand{\MRhref}[2]{%
  \href{http://www.ams.org/mathscinet-getitem?mr=#1}{#2}
}
\providecommand{\href}[2]{#2}

\end{document}